\documentclass[prl,aps,floats,superscriptaddress]{revtex4}
\usepackage{epsfig}
\usepackage[all]{xy}
\usepackage{graphicx}

\newcommand{\be}{\begin{equation}}
\newcommand{\ee}{\end{equation}}
\newcommand{\ba}{\begin{eqnarray}}
\newcommand{\ea}{\end{eqnarray}}
\newcommand{\bea}{\begin{eqnarray}}
\newcommand{\eea}{\end{eqnarray}}
\newcommand{\bean}{\begin{eqnarray*}}
\newcommand{\eean}{\end{eqnarray*}}
\newcommand{\bml}{\begin{mathletters}}
\newcommand{\eml}{\end{mathletters}}

\def\ie{{\it i.e. }}

\def\n{{\mathrm{n}}}
\def\L{{\cal{L}}}
\def\k{{\cal{K}}}
\def\R{{\cal{R}}}

\def\pp{/ \hspace {-2pt} /}

\begin{document}

\title{Einstein gravity on an even codimension brane}
\author{Christos Charmousis}
\affiliation{LPT, Universit\'e de Paris-Sud, B\^at. 
210, 91405 Orsay CEDEX, France}
\author{Robin Zegers}
\affiliation{LPT, Universit\'e de Paris-Sud, B\^at. 
210, 91405 Orsay CEDEX, France}
\affiliation{APC, Universit\'e Paris VII, 2 place Jussieu - 75251 Paris Cedex 05, France}
\email{christos.charmousis@th.u-psud.fr}
\email{robin.zegers@th.u-psud.fr}
\preprint{LPT-05-10}
\begin{abstract}
We give the equations of motion for a self-gravitating Dirac p-brane
embedded in an even co-dimension spacetime. The dynamics of the bulk 
are dictated by Lovelock gravity and permit matching conditions, even when the
codimension is strictly greater than 2. We show that the equations of motion
involve both
induced Lovelock densities on the brane and  regular 
extrinsic curvature terms. The brane dynamics can be derived from an
exact (p+1)-dimensional induced action. 
The Dirac charge is carried by an overall (solid) angle
defect which sets the Planck scale on the brane. 
In particular, if the codimension is greater than the worldvolume dimension of
the p-brane, we show that the extrinsic curvature terms cancel,
leaving an exact induced Lovelock theory. For example, a 3-brane embedded in
8 or 10 dimensions obeys Einstein's 
equations with a cosmological constant. 
\end{abstract}

\maketitle

In this letter, we consider the problem of finding matching conditions and
therefore the dynamics of an even
codimension self-gravitating brane. 
Consider a p-brane $\Sigma$ embedded in
$D=p+1+2n$ dimensions and suppose that $\Sigma$   
carries some 
 localised 
energy-momentum tensor~{\footnote{Greek letters run through brane coordinates
    while capital Latin from the begining of the alphabet through bulk coordinates}},
\be
\label{set}
T_{AB}= \left ( \begin{array}{cc}
S_{\mu \nu} & 0 \\
0&0
\end{array} \right ) \delta_{\Sigma} \, .
\ee
The Dirac distribution $\delta_\Sigma$ on $\Sigma$ 
signifies that the brane is of zero thickness. The question we address is:
What are the equations of motion for this self-gravitating p-brane  sourced by
the distributional energy momentum tensor (\ref{set})?  
For codimension 1, the answer is given by the well-known 
Israel junction conditions \cite{Israel} where, if the induced 
metric is continuous,  a discontinuity in the first 
derivatives of the
metric accounts for the Dirac charge in (\ref{set}). 
If the codimension
is strictly greater than 2 then there are no distributional 
matching conditions, at least in Einstein
gravity. Thus, finite thickness is needed in order to 
obtain non-trivial self-gravitating equations
of motion for $\Sigma$. Even when the codimension is equal to 2 (for $D=4$ say), one only knows the
self-gravitating field of a straight cosmic string, \ie one induced by a pure
tension matter tensor, which gives an overall conical deficit angle \cite{Vilenkin}. 

In this letter, we show that in order for
the bulk field equations to carry the necessary Dirac distribution, one has to
consider an extension of Einstein's theory to higher dimensions, that is
Lovelock theory. 
We demonstrate that this extension of the gravity theory, not only permits
matching conditions at the distributional level, 
but furthermore that the induced equations of 
motion for the p-brane are just those of ordinary 4 dimensional gravity. 

The main characteristic of Lovelock theory \cite{lovelock} is that it is 
identical to Einstein's theory in 4 dimensions, while in higher dimensions, it gives the most general gravity theory with  
second order field equations, conservation of energy and 
ghost-free perturbations \cite{Zwei}. In other words Lovelock theory obeys all
of Einstein's postulates. Lovelock's theory is special in its intimate
relation to generalised Euler characteristics which provide a topological
characterisation of manifolds. The most elegant way
to see this is by using differential forms.  

Consider a $D$ dimensional 
spacetime of metric $g=\eta_{AB} \theta^A \otimes \theta^B$,
where $\theta^A$ 
is the dual basis to the orthonormal local 
frame $\theta^A(e_B) = \delta^A{}_B$.
Curvature of spacetime is described by the curvature 2-form,
\be
\label{curv}
\R^A{}_{B}=\frac{1}{2}R^A{}_{BCD} \theta^C \wedge \theta^D \, ,
\ee
where $R^A{}_{BCD}$ are the Riemann tensor components. It is useful to
consider the $(D-k)$-forms,
\be
\theta^{\star}_{A_1 \cdots A_k}=\frac{1}{(D-k)!} \epsilon_{A_1 \cdots A_k
A_{k+1} \cdots A_D} \theta^{A_{k+1}} \wedge \cdots \wedge \theta^{A_D} \, ,
\ee
with $\epsilon_{A_1 \cdots \cdots A_D}$ the totally antisymmetric
pseudotensor. Note that $\theta^{\star}$ is the volume element. The useful
identity, 
\be
\label{ident}
\theta^B\wedge
\theta^{\star}_{A_1...A_k}=\delta^B_{A_k}\theta^{\star}_{A_1...A_{k-1}}
-\delta^B_{A_{k-1}}\theta^{\star}_{A_1...A_{k-2}A_k}+(-1)^{k-1}\delta^B_{A_{1}}\theta^{\star}_{A_2...A_k}
\ee
enables one to go back to familiar component language.
The Lovelock langrangian is a sum of densities of the form,
\be
\label{z}
{\cal L}_{(k)}=\R^{A_1 B_1}\wedge \dots\wedge \R^{A_k B_k} \wedge
\theta^{\star}_{A_1B_1 \dots A_kB_k}= \bigwedge_{i=1}^k \R^{A_i B_i}
\wedge \theta^{\star}_{A_1B_1...A_kB_k} \, ,
\ee
where $k$, called the rank, denotes the power of the curvature 2-form and we
have $k<{D\over 2}$. For $k=0,1,2$, for example, 
we have respectively the volume element, Ricci and Gauss-Bonnet scalars as can
be seen by applying (\ref{ident}). 
If spacetime is even dimensional then for $k=D/2$, $\L_{D/2}$ 
is the Euler density whose integral over
a compact manifold $M_D$ {\footnote{If spacetime is not compact one adds the relevant
  boundary term \cite{Chern}. See also \cite{GH}, \cite{Myers}}} yields its
Euler characteristic $\chi[M_D]$ according to 
\be
\chi \left [ M_{D} \right ]= \frac{1}{(4\pi)^{D/2} (D/2)!} \int_{M_{D}} {\cal
  L}_{(D/2)} \, .
\ee
This is the Chern-Gauss-Bonnet formula which says that the integral depends on
the topology  and not on the dynamics of $M_{D}$. 
Therefore in $D=4$, the Gauss-Bonnet term, $k=2$, does not contribute to the
field equations which can equally be derived from the usual Einstein-Hilbert action with a cosmological constant. 
In other words, each
topological invariant $\L_{(k)}$ 
becomes dynamical in one higher dimension, giving an extra term in the
Lovelock action. 
The field equations
are obtained by a variation with respect to the frame,
\be
\label{ll}
\sum_{k=0}^{[(D-1)/2]} \alpha_k {\cal E}_{(k) A} = -2T_{AB} \theta^{{\star} B}
\, ,
\ee
where $[\mbox{ }]$ stand for the integer part, and ${\cal E}_{(k) A}$ is the k-th rank (D-1)-form,
\be
\label{klov}
{\cal E}_{(k) A}= \bigwedge_{i=1}^k \R^{A_i B_i} \wedge 
\theta^{\star}_{AA_1B_1 \dots A_kB_k} \, .
\ee

Now, focus on the  $p$-brane $\Sigma$ of even
co-dimension $2n$. Let $e_\mu$ be the
$p+1$ unit vectors that are everywhere tangent to the brane, $\n_I$ be the
$2n$ unit vectors that are everywhere normal to $\Sigma$ with the label $N$
denoting the radial normal vector. Similarily, we split  $\theta^A$ into
tangent 1-forms $\theta^\mu$ and normal 1-forms $\theta^I$. One can then
deduce the first and second  fundamental forms of the brane, respectively; 
\ba
h &=& \eta_{\mu\nu} \theta^\mu \otimes \theta^\nu \\
K_{I\mu\nu} &=& g(\nabla_{e_\mu} \n_I, e_\nu) 
\ea 
the induced metric and extrinsic curvature.
The Gauss-Codazzi equations can also be written in form formalism. 
The  parallel projection along the brane coordinates gives the Gauss equation,
\be
\label{gform}
\R^\mu_{{\pp}\nu}\equiv \frac{1}{2}R^\mu{}_{\nu\lambda\rho} \theta^\lambda \wedge \theta^\rho= \Omega^{\mu}_{\pp \nu}- \k^{\mu}_{{\pp}I} \wedge
\k^I_{{\pp}\nu} \, ,
\ee
where 
$$\k^I_{\pp\mu}= K_{\nu \mu}^I  \theta^{\nu} \, ,$$
is the extrinsic curvature 1-form and $\Omega^\mu_{{\pp}\nu}$ is the {\it induced curvature
2-form} of the brane, associated with the induced metric $h_{\mu\nu}$.

The equations of motion for $\Sigma$ can now be derived; see \cite{CZ} for
details.
The main idea is the following. 
Given the form of (\ref{set}), we look for geometric terms in the
$\mu\nu$-components  on the LHS of (\ref{ll}) that can carry a Dirac
charge. These can  be equated with the brane
energy momentum tensor. This is exactly what one does for hypersurfaces, the
difficulty here being that like cosmic strings, the normal sections to the brane also have some
non-trivial geometry. So the first step is to define locally a normal section,
$\Sigma_\perp$, at each point of the brane, with specific regularity properties. We then
integrate the equations of motion over an arbitrary such section 
$\Sigma_\perp$. Since the Dirac distribution is insensitive to the
microscopic features of $\Sigma$, the only terms
that can contribute are those independent of geometry  deformations over 
$\Sigma_\perp$. We prove that the only terms in (\ref{ll}) having this property are locally exact forms on 
$\Sigma_\perp$. This is similar to the approach of Chern \cite{Chern} for the proof
of the extension of the  Gauss-Bonnet theorem, however applied to
$\Sigma_\perp$ rather than to the
whole manifold. The charged forms are precicely the Chern-Simons forms
of $S_{2n-1}$ which are pulled back by the radial normal vector field $N$ to $\partial
\Sigma_\perp$. When integrated out over $\Sigma_\perp$, they yield 
the charge or topological defect of $(1-\beta) Area(S_{2n-1})$,
exactly in the same way as the defect angle for the infinitesimal cosmic string
in 4 dimensions. Due to their topological character, we call these 
 topological matching conditions. 
They read \cite{CZ},
\be
\label{topo}
(1-\beta) \mbox{Area}({S}_{2n-1}) \sum_{\tilde{k}=0}^{\left [{p\over 2}\right ]} 
\tilde{\alpha}_{{k}} \sigma^{{\pp}}_{(n,\tilde{k})\mu}(P) = -2 S_\mu^\nu (P)
\theta^{\star {\pp}}_\nu \, ,
\ee
where we have set $\tilde{\alpha_{{k}}}=2^{2n-1}k!(n-1)!\alpha_k/(k-n)!$ and the
smooth parallel forms $\sigma^{{\pp}}_{(n,k)\mu}$ are given by
\be
\label{lemon}
\sigma^{{\pp}}_{(n,k)\mu}= \sum_{j=0}^{{\mbox{\footnotesize{min}}\left (n-1,
\tilde{k} \right)}}  \left( ^{\,\tilde{k}}_{j} \right )
\left (\bigwedge_{l=1}^{\tilde{k}-j} \R^{\lambda_{l} \nu_{l}}_{{\pp}} \right ) \wedge
\left ( \bigwedge_{l=1}^{2j} \k^{\rho_l}_{{\pp} N} \right ) \wedge
\theta^{\star {\pp}}_{\mu \lambda_{1} \nu_{1} \cdots \lambda_{\tilde{k} - j} \nu_{\tilde{k} - j}
  \rho_{1} \cdots \rho_{2j}} \, .
\ee
Finally, $\tilde{k}=k-n$, which ranges between $0\leq \tilde{k}\leq \left
  [{p\over 2}\right ]$, will turn out to be the induced Lovelock
rank of (\ref{lemon}). The parallel forms (\ref{lemon}) will dictate the
dynamics of the brane. Their expressions involve powers of the
projected Riemann tensor $\R_{\pp}$ on the brane and even powers of
the radial extrinsic curvature $\k_N$  of $\Sigma$. To see this, consider for
each $\tilde{k}$ the first term in the sum, $j=0$, which reads,
\be
\label{marion}
\sigma_{(\tilde{k},0)\mu}^{\pp}=\left (\bigwedge_{l=1}^{\tilde{k}} \R^{\lambda_{l}
    \nu_{l}}_{{\pp}} \right )  \wedge \theta^{\star {\pp}}_{\mu \lambda_1
    \nu_{1}...\lambda_{\tilde{k}}\nu_{\tilde{k}}} \, .
\ee
Note the similarity to the bulk Lovelock densities, (\ref{klov}). 
Clearly, using (\ref{gform}), we see the appearance of induced Lovelock densities involving $\Omega_{\pp}$ accompanied by even powers of the extrinsic curvature $\k_N$. The former describe the induced quantites of the brane and the latter how $\Sigma$ is embedded in the bulk. In particular, for $\tilde{k}=0$ we will have a pure tension term, for $\tilde{k}=1$ an Einstein term and for $\tilde{k}=2$ a Gauss-Bonnet term with extrinsic curvature terms. Note that the highest rank on the brane originates from the highest rank Lovelock term in the bulk.  This agrees with the fact that in Einstein gravity, $k=1$, there are no matching conditions beyond $n>2$.

Let us compare our result with the work of \cite{Greg}. Thus, let us consider
the case of a 3-brane embedded in 6 dimensional spacetime, for which there are
only two terms in (\ref{lemon}),
\ba
\label{einscodim2}
\sigma_{(1,0) \mu}^{\pp} &=&  \theta^{\star {\pp}}_\mu, \qquad \sigma_{(1,1)
  \mu}^{\pp} = \R_{{\pp}}^{\nu \rho} \wedge \theta^{\star
{\pp}}_{\mu \nu \rho} \, . 
\ea
Thus, using (\ref{ident}), (\ref{curv}) and (\ref{gform}), we obtain from
  (\ref{topo}), 
\be
2 \pi \, (1-\beta)  \left \{ -\alpha_1 h_{\mu\nu}
 + 4 \alpha_2  G_{\mu\nu}^{(ind)}  - 4 \, \alpha_2
\,  W_{\mu\nu} \right \} 
=  S_{\mu\nu}  \, ,
\ee
where \cite{Greg}
\be
\label{w}
W_{\mu}^{\lambda}=K_{N} K^{\lambda}_{\mu N}  - K^\nu_{\mu N}
K^{\lambda}_{\nu N}  - \frac{1}{2}
\delta^{\lambda}{}_{\mu} \left( K_{N}^2 - K^\nu_{\rho N} K^{\rho}_{\nu N}
\right ) \, .
\ee
Note that the bulk Einstein term, $k=1$, 
only allows for an effective cosmological constant on the 
brane, while the
Gauss-Bonnet term, $k=2$, induces the Einstein tensor  
for the brane's equation of motion. This equation is  similar to the one
found in \cite{Greg}, the difference being that here,
the extrinsic geometry is supposed perfectly
regular. Indeed mathematically  there is no reason to suppose that the
extrinsic curvature has a jump, the topological defect carrying the necessary
charge. Aditionally, these matching conditions provide the maximal regularity
for the bulk metric. Furthermore, since the induced and extrinsic geometry are
smooth, the equations of motion are seen to actually originate from a simple
action taken over $\Sigma$. In other words, the degrees of freedom
associated to the normal section are completely integrated out giving an exact
action for the brane's motion. In differential form language, it is straightforward to read off the langrangian density
in question. Literally, taking out the free index for the charge in
(\ref{einscodim2}) and using Gauss equation (\ref{gform}), we get
\bea
\label{melina}
S_{\Sigma}^{(p=3,n=1)} &=& 2 \pi (1-\beta) \, \int_{\Sigma} \left( \tilde{\alpha_1} \theta^{\star \pp}+ \tilde{\alpha_2} (\R^{\nu
\rho}_{{\pp}} \wedge \theta^{\star {\pp}}_{ \nu\rho} )\right)+\int_{\Sigma}
\L_{matter} \nonumber\\ &=& 2 \pi (1-\beta) \int_{\Sigma} \sqrt{-h} \, \left (\tilde{\alpha_1}+ \tilde{\alpha_2}
(R^{ind}-K^2+ K_{\mu\nu}^2) \right ) +\int_{\Sigma}
\L_{matter} \, ,
\eea
the cosmological constant plus Einstein-Hilbert action with an extrinsic
curvature term quite similar to the finite width corrections one obtains for
cosmic strings in flat spacetime \cite{bonjour}.
In other words, the only bulk quantity entering in the equations of motion is
the topological defect fixing the overall mass scale 
and the extrinsic curvature of the surface giving a matter like component in
the action.

Equation (\ref{topo}) has another surprising property. Indeed,  an important
simplification takes place if we suppose that 
\be
\label{isa}
\left[ p/2\right] +1 \leq n \, ,
\ee 
{\it ie} that the codimension of the brane is larger than its 
intrinsic dimension. In that case, using Gauss's equation, we have
\be
\label{simple}
\sigma^{{\pp}}_{(n,k)\mu}= \left (\bigwedge_{l=0}^{\tilde{k}} 
\Omega^{\lambda_{l} \nu_{l}}_{{\pp}} \right ) \wedge 
\theta^{\star {\pp}}_{\mu \lambda_1 \nu_{1} \cdots \lambda_{\tilde{k}} \nu_{\tilde{k}}} 
\ee
and the matching conditions (\ref{topo}) are simply the induced Lovelock
equations on the brane with no extrinsic curvature terms! Therefore, the action
for a distributional 3-brane embedded in $D=8,10,... 2d$ dimensions is exactly
the Einstein-Hilbert plus cosmological constant action
$$
S_{\Sigma}^{(p=3,n>1)}= (1-\beta) \mbox{Area}(S_{2n-1}) \int_{\Sigma} \sqrt{-h} \left
  ( \tilde{\alpha}_n+ \tilde{\alpha}_{n+1}
R^{ind} \right )+\int_{\Sigma} \L_{matter} \, ,
$$
with Planck scale set by
$$
M_{Pl}^2=(1-\beta)\, \mbox{Area}(S_{2n-1})\, \tilde{\alpha}_{n+1} \, .
$$
For a 4 or 5-brane, we will have in addition the Gauss-Bonnet term etc. In other
words, if the codimension verifies (\ref{isa}), all the extrinsic curvature
corrections drop out and there is a complete Lovelock reduction from the bulk
to the brane.   All the degrees of freedom originating from the bulk at zero
thickness level are exactly integrated out giving the most general classical equations of motion for the brane. 
As a consequence we have in particular energy-conservation on
the brane (see also \cite{kofinas}).

The Lovelock reduction presented here is valid only for even codimension
p-branes and this is clearly due to the presence of Euler densities for even
dimensional manifolds. When the codimension is odd, as for hypersurfaces, one
obtains \cite{CZ} for $\sigma_{\pp}$ (\ref{lemon}) odd
powers of $\k_N$, schematically, $\k_N$, $\k_N \R_{\pp}$, $\k_{N}^3$ etc. These
are clearly Chern-Simons forms on $\Sigma$ as one obtains for boundary terms,
\cite{GH}, \cite{Myers}. 

In \cite{CZ}, we also discuss examples of matching conditions for maximally symmetric
p-branes. Let us
briefly illustrate here, in component language, the case of a flat 1-brane embedded in a 6 dimensional
spacetime.
The situation is quite analogous to the case of the cosmic string. We
find that in order for spacetime to be free of naked singularities, it must be
locally flat apart from an overall solid  angle defect taken out of the normal
3-sphere. The metric reads
\be
\label{ruth}
ds^2 =-dt^2 + dz^2 + dr^2 + L^2(r)
\left (\beta^{-2} \, d\theta^2 + \beta^{-2} \, \sin^2 \theta \, d\phi^2
+ \sin^2 \theta \, \sin^2 \phi \, d\psi^2 \right ) \, ,
\ee
where $L'=\beta$ for $r>0$. The catch is that the charge can only be provided
by the Gauss-Bonnet term since we have a 4-dimensional Dirac distribution
$\delta_\Sigma=\frac{\delta (r)}{2\pi^2 \beta^{-2} L^3}$ in (\ref{set}). The relevant Gauss-Bonnet tensor components read
\be
\label{roberto}
H^t{}_t = H^z{}_z \sim \frac{4}{L^3}  \hat{f '} + \frac{4\delta(r)}{L^3}
\left [ f \right ] \, ,
\ee  
where the function
\be
\label{Lcone}
f(r) = \left \{ \begin{array}{cc}
(3\beta^2-L'^2)L' & \mbox{if $r > 0$} \\
3\beta^2-1 & \mbox{if $r=0$}
\end{array} \right .  
\ee
is discontinuous at the origin, giving the matching condition $\sigma=2\pi^2
\alpha_2 (1-\beta)^2(1+2\beta)/\beta^2$, where $\sigma$ is the tension. In fact the components
(\ref{roberto}) are excactly the Gauss-Bonnet scalar which when integrated
over the 4 dimensional normal section give according to the
Chern-Gauss-Bonnet formula the RHS of the matching condition. 

We should point out that the matching conditions (\ref{topo}) are necessary
but not sufficient conditions for the existence of a well defined Dirac
brane. One has further constraints originating from the remaining Lovelock
equations. These constraints,  among other things, condition the behaviour of the Riemann
tensor at the vicinity of the brane. In order for the matching conditions to
make any sense, the Riemann curvature has to be bounded, as in the above simple
example. Unlike the case for
hypersurfaces, where the position of the brane and the normal vector
permit 
the local reconstruction of spacetime, as 
codimension gets higher, spacetime dynamics seems to influence less and less the intrinsic dynamics
of $\Sigma$ and vice-versa. It seems for example, that an observer on a
3-brane, embedded in a  10 dimensional spacetime, will have no way of telling
(in the zero thickness limit) that she were part of a higher dimensional spacetime. To make sure that this is an acurate
predicition rather than a neat geometrical property of Lovelock gravity, one should consider small perturbations around (\ref{ruth}) and
see indeed that 4-D gravity is localised with the predicted volume element and no
scalar degrees of freedom  are present in the gravitational spectrum
{\footnote{Given the arbitrariness of the parameter $\beta$ one might expect the presence of a massless scalar associated
to the conical defect}}. This
is currently under investigation.

\noindent
It is a great pleasure to thank Roberto Emparan and Daniele Steer 
for discussions and helpful comments.


\begin{thebibliography}{99}

\bibitem{lovelock}D. Lovelock, J.Math.Phys. {\bf 12} (1971) 498.

\bibitem{Israel}
W.~Israel,
Nuovo Cim.\ B {\bf 44S10} (1966) 1
[Erratum-ibid.\ B {\bf 48} (1967\ NUCIA,B44,1.1966) 463].



\bibitem{Vilenkin}
A.~Vilenkin,
Phys.\ Rev.\ D {\bf 23}, 852 (1981).

\bibitem{Zwei}
B.~Zwiebach,
Phys.\ Lett.\ B {\bf 156} (1985) 315.


\bibitem{Chern} S.-S.~Chern, 
Ann.~Math.~45 (1944) 747-752. \\
S.-S.~Chern, 
Ann.~Math.~46 (1945) 674-684.

\bibitem{kofinas}
G.~Kofinas,
arXiv:hep-th/0412299.


\bibitem{GH}
G.~W.~Gibbons and S.~W.~Hawking,
Phys.\ Rev.\ D {\bf 15}, 2752 (1977).

\bibitem{Myers}
R.~C.~Myers,
Phys.\ Rev.\ D {\bf 36}, 392 (1987).

\bibitem{Greg} 
P.~Bostock, R.~Gregory, I.~Navarro and J.~Santiago,
Phys.\ Rev.\ Lett.\  {\bf 92}, 221601 (2004)
[arXiv:hep-th/0311074].



\bibitem{bonjour}
M.~Anderson, F.~Bonjour, R.~Gregory and J.~Stewart,
Phys.\ Rev.\ D {\bf 56}, 8014 (1997)
[arXiv:hep-ph/9707324].



\bibitem{CZ} C.~Charmousis and R.~Zegers, [arXiv:hep-th/0502170].


\end{thebibliography}
\end{document}